\documentstyle{./neu}                          
\input psfig.sty
\begin{document}
\title{%
GENERAL RELATIVISTIC POLAR \newline
PARTICLE ACCELERATION AND \newline
PULSAR DEATH}

\author{Jonathan ARONS \\
{\it Department of Astronomy, Department of Physics and  \newline
Theoretical Astrophysics Center, University of California at \newline
Berkeley, 601 Campbell Hall, Berkeley, CA 94720-3411, USA \newline 
arons@astroplasma.berkeley.edu} }

\maketitle

\section*{Abstract}
I summarize the theory of acceleration of non-neutral particle 
beams by starvation electric fields along the polar magnetic field
lines of rotation powered pulsars, including the effect of dragging
of inertial frames which dominates the acceleration of a space charge
limited beam. I apply these acceleration results to a new calculation
of the radio pulsar death line, under the hypotheses that pulsar
``death'' corresponds to cessation of pair creation over the 
magnetic poles {\it and} that the magnetic field has a locally
dipolar topology.  While the frame dragging effect in star centered
dipole geometry does improve
comparison of the theory with observation, an unacceptably
large fraction of the observed stars outside the bounds of pair
creation theory still persists. Offsetting the dipole improves the correspondence
between theory and observation. The result is a ``death valley'' for pulsars; 
acceptable comparison of observation and theory occurs if the
boundary of death valley corresponds to offsets of the dipole center
from the stellar center $\sim (0.7-0.8) R_*$. I also point out that pulsars
are absent for magnetic moments corresponding to star centered polar fields
in excess of $\sim 4 \times 10^{13}$ Gauss,
and I suggest that this absence is due to pairs forming as bound
positronium atoms in such strong fields, creating a neutral,
relativistically outflowing gas which cannot participate in
low altitude collective radio emission processes in such strongly
magnetized objects.

\section{Introduction}
Most of the scientific community which has an interest in the
physics of neutron stars believes that 
radio emission from Rotation Powered Pulsars (RPPs)
has its origin in the relativistic outflow
of electron-positron pairs along the polar magnetic field lines of
a dipole magnetic field frozen into the rotating neutron star (e.g., 
Arons 1992, Meszaros 1992). 

The evidence for dipole magnetic fields
in RPPs (and in any other neutron star) is indirect, coming primarily
from the electromagnetic theory of RPP spindown. The observed 
increasing pulse periods are readily explained using standard 
theoretical moments of inertia plus order of magnitude estimates,
derivable from dimensional analysis (Dyson 1971, Arons 1979, 1992),
of rotational energy loss driven by relativistic electromagnetic 
spindown torques,
\begin{equation}
\dot{E}_R = k \frac{\mu^2 \Omega_*^4}{c^3} = 
   -I \Omega_* \dot{\Omega}_*.
\label{rotloss}
\end{equation}
Here $\mu$ is the magnetic moment, $\Omega_*$ is the stellar angular
velocity with respect to inertial space far from the star, and $k$ is a
function of any other parameters of significance, with magnitude on
the order of unity.  In the vacuum theory (Deutsch 1955), 
$k =(2/3) \sin^2 i$, with $i$ the angle between the magnetic moment
and the angular velocity. Theoretical work on the
torques due to conduction currents steming back to Goldreich and Julian
(1969), coupled to the observation that spindown rates appear to
be independent of observationally estimated values of $i$ (Lyne and
Manchester 1988), suggest that in reality $k$ does not substantially
depend on $i$. In the subsequent discussion, 
I assume k = 4/9, the average of the vacuum value over the sphere. 
Application of
(\ref{rotloss}) to the observations of RPPs' periods
($P = 2 \pi /\Omega_*$ and period derivatives
($\dot{P} = -2 \pi \dot{\Omega}_* / \Omega_*^2$)
yields $\mu \sim 10^{30}$ cgs for ``normal''
RPPs, and $\mu \sim 10^{27}$ cgs for millisecond RPPs. These
results are reasonably firm, the main uncertainty coming from
the derived values of $\mu$ being proportional to $k^{-1/2}$.

The electromagnetic torque interpretation of pulsar spindown
constrains only the exterior dipole moment of the magnetic field.
However, not long ago Rankin (1990) presented 
strong evidence in favor of a low altitude ($r \approx R_*$) dipole 
geometry for the site of the core component of pulsar radio emission.
Arons (1993) gave evidence that spun up millisecond
pulsars must have a substantially dipolar large scale field 
at low altitute.

Electron-positron pair creation at low altitude above the polar caps 
has long been hypothesized to be an essential ingredient of pulsar 
radio emission, starting with Sturrock's (1971) pioneering work.  
If so, all observed pulsars must lie in the region of 
$P-\dot{P}$ space where polar cap acceleration has sufficient vigor 
to lead to copious pair production. Yet, to date, all 
{\it internally consistent} theories of polar cap pair 
creation have required hypothesizing a large scale ({\it e.g.}, quadrupole) 
component of the magnetic field with strength comparable to that of the
dipole (Ruderman and Sutherland 1975, Arons and
Scharlemann 1979, Barnard and Arons 1982, Gurevich and Istomin 1985).
These non-dipole components were invoked in order to increase the opacity
of the magnetic field to pair creating gamma rays.  Non-dipole low altitude 
fields can have magnetic radii of curvature on the order of $R_*$ or less,
a factor of 50-100 smaller than the radii of curvature of star centered dipole
field lines near the magnetic poles. The resulting increase of optical depth
allowed the pair creation models
to cover the whole $P, \; \dot{P}$ diagram. However, such strong magnetic 
anomalies contradict the evidence in favor of an apparently
dipolar low altitude geometry; the alteration of the magnetic geometry
also ruins the internal consistency of many models' electrodynamics. 

Both early (Sturrock 1971) and more
recent work on polar cap electrodynamics and its implications for
the occurrence of pair creation in $P, \; \dot{P}$ space either employ
incomplete (e.g. Sturner {\it et al.} 1995) or erroneous (Sturrock 1971, Mestel 
and Shibata 1994, Bjornsson 1996) theories of polar cap particle 
acceleration. Most of
the internally consistent theories also violate other 
observational constraints, especially with regard to polar
cap heating (Arons 1992), which creates pulsed thermal X-ray emission from hot spots in
excess of what is seen (Becker and Tr\"{u}mper 1997, Pavlov and Zavlin 1997).  
While the Arons and Scharlemann (1979) model does not have this problem,  
in star centered dipole 
geometry it dramatically fails to account for pulsar emission over most
of the $P-\dot{P}$ diagram and predicts radio polarization variations in 
contradiction to the observations (Narayan and Vivekanand 1982).

Here I describe a low altitude polar cap
acceleration theory which successfully associates pulsar ``death'' with
the cessation of pair creation in an {\it offset} dipolar low altitude 
magnetic field.  The basic acceleration physics is that
of a space charge limited relativistic particle beam accelerated 
along the field lines by the starvation electric field,
as in the Arons and Scharlemann theory, but with the additional
effect of inertial frame dragging, first pointed out by Muslimov and Tsygan 
(1990, 1992) and by Beskin (1990).  

This effect
causes the accelerating electric field to be about an order of magnitude
larger than that calculated by Arons and Scharlemann for pulsars
near the death line, which substantially improves the size of
the region in $P, \; \dot{P}$ space in which polar cap pair creation
occurs, but still does not allow the theory to fully account for
the observed pulsar distribution, in star centered dipole geometry.
If the dipole's center is offset from the stellar center along a vector parallel
to the dipole moment itself, an offset which automatically preserves the
symmetries built into the highly successful Radhakrishnan and Cooke (1969)
model of polarization swings, the magnetic field at one pole becomes substantially
stronger than it would be if the same magnetic dipole were star centered.
If the offset is substantial (as much as 80\% of the stellar radius turns out to
be required), all pulsars can be accommodated within a single pair creation
theory.  The location of an individual pulsar's pair creation death depends
on the magnitude of the offset, thus yielding a ``death valley'' (Chen and
Ruderman 1993) for the whole pulsar population.

\section{Polar Acceleration}

Prior to the work of Muslimov and Tsygan and of Beskin, study of polar cap
relativstic particle acceleration in the 1970's had led to the 
conclusion that acceleration of a space charge limited particle beam from
the stellar surface with energy/particle high enenough to emit
magnetically convertible curvature gamma rays occurs because of curvature 
of the magnetic field (Scharlemann {\it et al.} 1978, Arons and Scharlemann
1979). With field line curvature, matching of the beam 
density to $\eta_R$ occurs only at the surface. Along field lines 
which curve toward the rotation axis (``favorably curved'' field lines, 
$|\eta_{\rm beam} /\eta_R| < 1$), the beam fails to short out the vacuum, with 
$|(\eta_{\rm beam}  -\eta_R )/\eta_R| \sim R_* /\rho_B $. 
$R_* =10 R_{10}$ km is the stellar
radius and $\rho_B$ is the radius of curvature of the magnetic field lines. 
Therefore, particles accelerate along $B$ through a potential drop
\begin{equation}
\Delta \Phi_\parallel \approx \Phi_{\rm pole} 
    \left(  \frac{R_*}{\rho_B} \right) 
    \sim 10^{-2} P^{-1/2} \Phi_{\rm pole},
\label{eq:SAF-pot}
\end{equation}
where $P$ is the rotation period in seconds, 
and the numerical value assumes field lines have dipolar radius of curvature. 
Here
\begin{equation}
\Phi_{\rm pole} \equiv \frac{\Omega_*^2 \mu}{c^2} = 
    1.09 \times 10^{13} \left( \frac{I_{45}}{k} \right)^{1/2}
      \left(\frac{\dot{P}_{15}}{P^3} \right)^{1/2} \; {\rm Volts},
\label{eq:polar-pot}
\end{equation}
with $\dot{P}_{15} \equiv \dot{P} /10^{-15} \; {\rm s/s}$ and 
$I_{45} = I/10^{45}$ g-cm$^2$. Particles drop through the potential 
(\ref{eq:SAF-pot}) over a length $L_\parallel \sim R_*$ (an electric field
of magnitude $\sim 10^7 -10^8$ Volts/meter, for normal {\it and} millisecond
pulsars). Note that $\Phi_{pole}$ is proportional to the magnetic flux contained 
in the tube of open field lines. Therefore, the total potential of a pole is
independent of the magnetic topology, {\it if, and only if, the open field
lines map onto a single, more or less round polar cap}. 

Curvature gamma rays have typical energy 
$\varepsilon_c \sim (\hbar c / \rho_B ) (e \Delta \Phi_\parallel / m c^2 )^3 
 \propto  \Phi_{\rm pole}^3 /\rho_B^4$, while the optical depth for pair
creation, due to one photon conversion of gamma rays emitted by electrons (or
positrons) accelerating through the unshorted potential (\ref{eq:SAF-pot}), 
can be shown to be (Arons and Scharlemann 1979, Luo 1996, Bjornsson 1996)
$\tau = \Lambda \exp[-a (mc^2 /\varepsilon_c) (B_q /B_*) (\rho_B /R_* )]$,
where $a$ is a pure number (typically  $ \sim 30$) and $\Lambda$ is a 
combination of the basic parameters which is quite large 
($\ln \Lambda \sim 20$). Pair creation does not short out the acceleration
if $\tau \ll 1$; thus, if pairs are important for radio emission, a reasonable
theoretical definition of the death line is $\tau = 1$. A more precise definition
requires estimating the number of charges that must be added to restore
the total charge density to $\eta_R$ and reduce $E_\parallel$ to zero. This
refinement is included in the results shown in Figures  \ref{fig:death-valley} and  
\ref{fig:phi-P}  Using 
$B_* = 2 (\Phi_{\rm pole} / R_*) (c / \Omega_* R_*)^2$, the potential
(\ref{eq:SAF-pot}) and setting $\tau$ equal
to unity yields the death line, expressed as $\Phi_{\rm death} (P)$
such that stars  $\Phi_{\rm pole} < \Phi_{\rm death}$ do not make pairs.
This death line, first found by Arons and Scharlemann (1979), appears as the
dashed line in
Figure \ref{fig:centered}, when $\rho_B$ assumes the star centered dipole value
$ \sim (R_* c /\Omega_*)^{1/2}$.  

The expression for $\tau$ shows that 
$\Phi_{\rm death} \propto \rho_B^{5/4}$.  Arons and Scharlemann argued, following 
Ruderman and Sutherland (1975), that if magnetic anomalies reduced $\rho_B$
to be on the order of $R_*$, better agreement with the cessation of radio
emission might be achieved, for ``normal'' pulsars, almost all of which
have periods between 0.1 and 1 second.  Figure \ref{fig:centered} shows clearly that the
large dynamic range in $\Phi_{pole}, \; P$ space made available by the
cataloging of millisecond pulsars falsifies
even this ``fudged'' version of the theory - the scaling with period,
$\Phi_{\rm death} \propto P^{3/8}$, flatly disagrees with the shape of
the boundary of pulsar radio emission in the $\Phi_{pole}, \; P$ diagram.
When combined with the more recent arguments in favor of dipolar topology 
for the low altitude
magnetic field, I drew the conclusion 
either that something else governs the
low altitude acceleration which leads to pair creation, or that pair creation is
not important to radio emission.

Muslimov and Tsygan (1990, 1992) revivified this subject by uncovering a
previously overlooked effect on the acceleration of the non-neutral beam from
the stellar surface. Stellar rotation drags the inertial frame into rotation,
at the angular velocity 
$\omega_{LT} = (2 G I/R_*^2 c^2) \Omega_* (R_* /r)^3$,
where I is the moment of inertia. Therefore, the
electric field required to bring a charged particle into corotation is
${\bf E}_{co}=-(1/c)[({\bf \Omega}_* - {\bf \omega}_{LT}) \times {\bf r}] \times {\bf B}$;
the rotation of the magnetic field with respect to {\it local} inertial space, not
inertial space at infinity, determines the electric field which in turn sets
a charged particle's ${\bf E \times B}$ drift velocity. The charge density required to
support this {\it local} corotation electric field therefore is 
\begin{equation}
\eta_R = - \frac{({\bf \Omega}_* - {\bf \omega}_{LT}) \cdot {\bf B}} {2 \pi c} =
  - \frac{{\bf \Omega}_* \cdot {\bf B}} {2 \pi c} 
    \left[ 1 - \kappa_g \left(\frac{R_*}{r^3} \right) \right],
\label{eq:LTcorotdens}
\end{equation}
where $\kappa_g = 2 G I /R_*^3 c^2 = 0.17 (I_{45} /R_{10}^3)$. Relativistic
space charge limited flow from the surface 
has a beam charge density 
$\eta_b  
 = - ({\bf \Omega}_* \cdot {\bf B}_* /2\pi c ) (1-\kappa_g) (B/B_*)$. Above the
surface, this charge density is too small to short out $E_\parallel$ on
{\it all} polar field lines, not just the favorably curved part of a polar
flux tube, thus restoring the possibility of polar cap acceleration models
being in accord with the observed rough symmetry of radio emission with respect to
the magnetic axis (e.g., Lyne and Manchester 1988). One can graphically
describe this general relativistic origin of electrical starvation simply as
the consequence of the field lines rotating faster with respect to inertial
space as the radius increases, at the angular speed 
$\Omega_* - \omega_{LT}(r) = \Omega_* [1 - \kappa_g (R_* /r)^3 ]$.
The constraint of relativstic flow along
$B$ allows the beam to provide only a charge density sufficient to support 
corotation at the angular speed $\Omega_* (1 - \kappa_g)$. The difference
not surprisingly leads to a accelerating potential drop 
\begin{equation}
\Delta \Phi_\parallel \approx \kappa_g \Phi_{pole} [1 - (R_*/r)^3].
\label{eq:polarpot}
\end{equation}
For normal
pulsars with dipole fields, $\kappa_g \sim 10 R_* /\rho_B \approx 10 P^{-1/2}$, so that the
effect of dragging of inertial frames on the beam's acceleration
can yield curvature gamma ray energies 1000 times 
greater than occur in the Arons and Scharlemann pair creation theory, for
normal pulsars; for MSPs, the theories yield comparable results, although
of course the symmetry of the beam with respect to the magnetic axis differs.
Expression (\ref{eq:polarpot}) applies
to the polar flux tube at altidudes  greater than the width of
the polar flux tube; closer to
the surface, fringe fields cause the potential to be smaller.

\section{Death Lines and Death Valley}

When curvature emission is the only source
of gamma rays, and curvature emission does not limit an accelerating 
particle's energy, the condition 
that the number of pairs created be just
such as to reduce both $E_\parallel$ and $\nabla \cdot {\bf E}_\parallel $
to zero (which is almost
the same as the optical depth unity condition) yields the death line for a star
centered dipole (Arons, in preparation):
\begin{equation}
\Phi_{death} = 1.9 \times 10^{13} 
   \left( \frac{R_{10}^4}{I_{45} \cos i} \right)^{3/4} P^{-1/4} \; {\rm Volts}.
\label{eq:cntrdcrvdth}
\end{equation}

\begin{figure}
\psfig{file=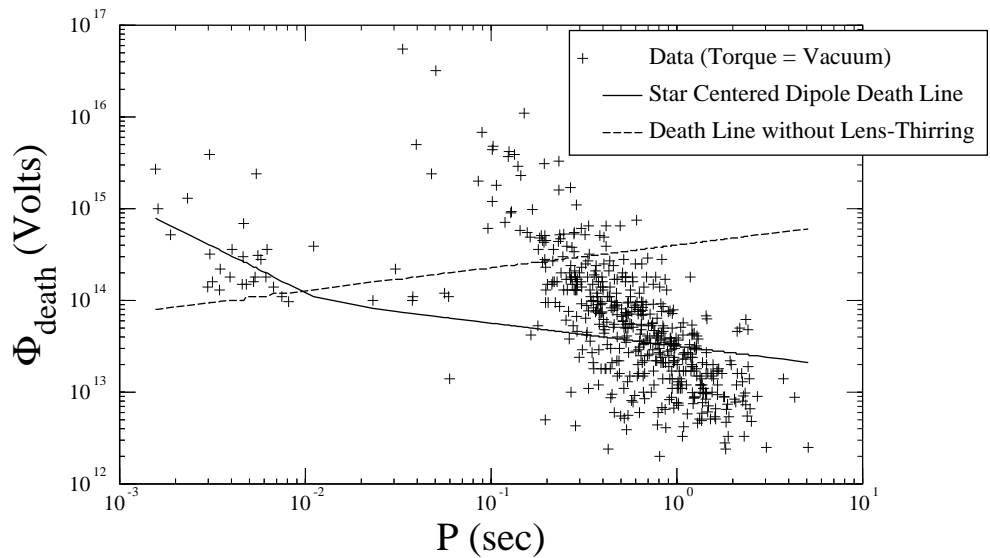}

\caption{Pair creation Death Lines for Star Centered Dipoles.  Solid line:
Standard gamma ray emission and absorption in a star centered dipole, when
the beam acceleration model incorporates the effect of inertial frame dragging.
Dashed line:  Same geometry, gamma ray and pair physics, but with inertial
frame dragging neglected in the particle acceleration theory.
\label{fig:centered}}
\end{figure}

\noindent This result appears in Figure \ref{fig:centered} The effects 
of curvature radiation reaction, important at short periods, are also included. 

Dragging of inertial frames clearly improves the agreement between the
boundary of pair activity in the $\Phi - P$ diagram and the region where
pulsars occur, but the discrepancy is still too large - something else
is missing.  If the field geometry must be {\it locally} dipolar at low altitude, 
then the only ingredients still not included are 1) offset of the dipole 
from the stellar center and 2) additional gamma ray emission and absorption 
processes.

\begin{figure}
\psfig{file=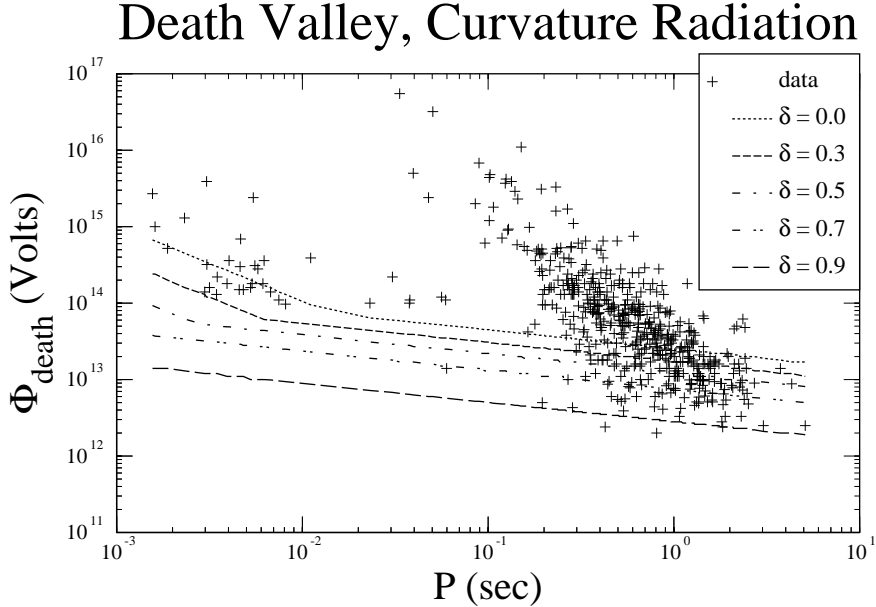}
\caption{Death Valley for offset dipoles, with magnetic moment ${\bf \mu}$
     parallel to the offset vector ${\bf \delta}$, assuming no inverse
     Compton gamma rays and one photon magnetic pair creation.  Radiation
     reaction significantly limits the particles' energies at high voltages 
     and short periods. \label{fig:death-valley-curv}}
\end{figure}

I discuss only the simplest dipole offset here, namely, when the magnetic
field is that of a point dipole, with the center of the dipole displaced
from the stellar center by an offset vector ${\bf \delta}$  parallel to 
${\bf \mu}$. This has the effect of increasing the
magnetic field at one pole to strength $B_* = 2 \mu / (R_* - \delta )^3$, with
a resulting drastic increase in the gamma ray opacity, while leaving the 
accelerating potential unaltered. The result is the death
valley shown in Figure \ref{fig:death-valley-curv} 
Clearly, dipole offsets
do allow the ``classical'' theory of pulsar death to survive modern observations,
although at the price of displacements of the dipole center from the stellar
center comparable to moving the dipole's center to the base of  the crust.

This estimate of death valley's extent assumes curvature emission and magnetic
conversion to be the only sources for gamma ray emission and absorption.
ROSAT observations have revealed the long sought thermal X-rays from neutron star
surfaces (Becker and Tr\"{u}mper 1997).  Resonant Compton scattering creates 
magnetically convertible gamma rays at a spatial rate 
$(dN_\gamma /ds)_{rC}  \propto T_* /\Gamma^2$,
(e.g., Luo 1996) where $T_*$ is the temperature of the cooling 
neutron star (polar cap heating is unimportant near the death line) and 
$\Gamma = e \Phi /m_\pm c^2$ is the Lorentz factor of an electron or
positron in the beam. Compton scattering thus can become a significant
source of gamma rays in stars with smaller accelerating potentials. 
In contrast, the spatial rate of curvature 
emission, $(dN_\gamma /ds)_c \propto \Gamma /\rho_B$, shows that curvature
emission dominates gamma ray emission for stars with large voltages.
Compton scattering thus may contribute significantly for stars with low 
overall voltage, just where the theory based solely on curvature emission encounters
the most trouble in accounting for the data. 

\begin{figure}
\psfig{file=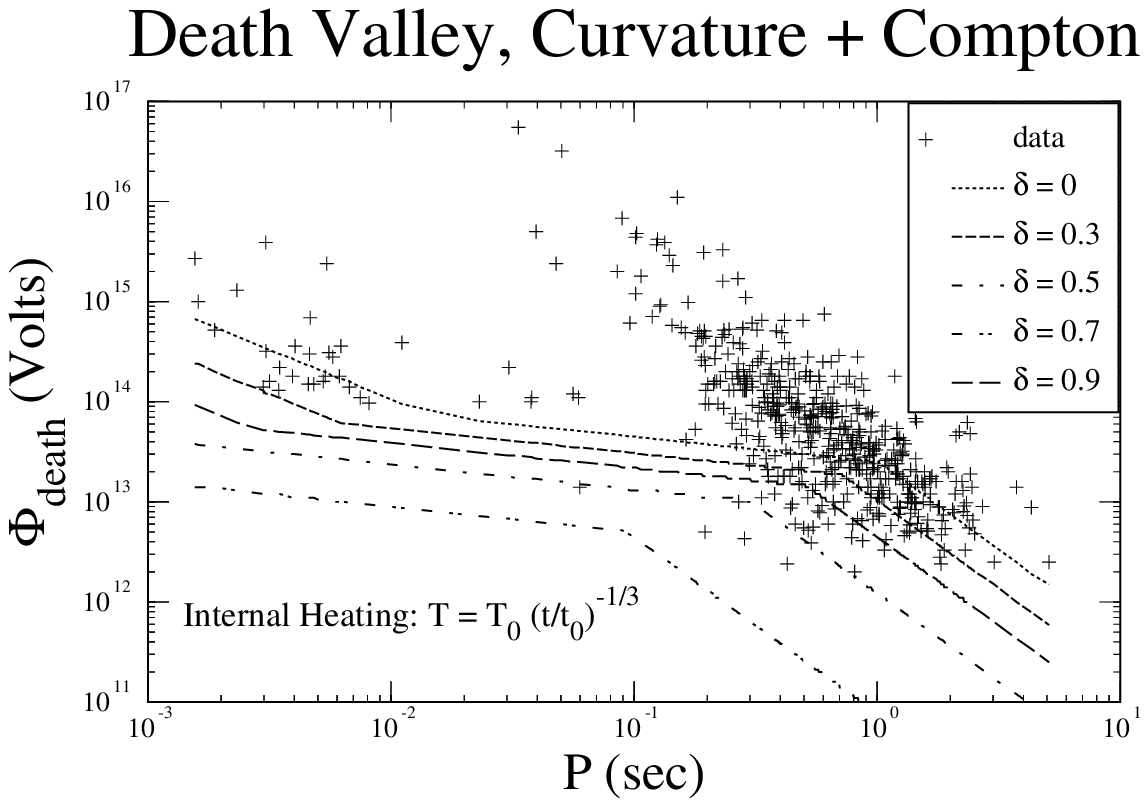}
\caption{Death Valley for an offset dipole parallel to the offset vector,
  with both curvature and {\it resonant} inverse Compton emission and stellar temperature
  kept high by internal heating.  \label{fig:death-valley}}
\end{figure}

Indeed, this expectation is correct,
{\it if} internal heating (e.g., Umeda {\it et al.} 1993) keeps the 
surface temperature above $10^5$ K to spindown ages in excess of $10^{7.5}$
years. In this case, resonant Compton scattering of thermal photons by a polar electron 
beam {\it does}
extend death valley to include all the observed pulsars, with somewhat less
drastic offsets required, as is shown in Figure \ref{fig:death-valley}

\begin{figure}
\psfig{file=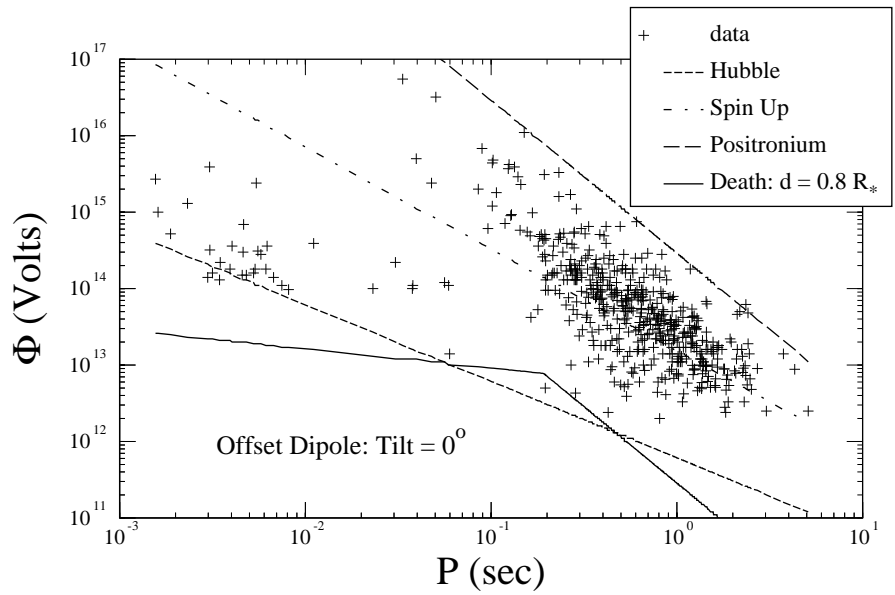}

\caption{The complete $\Phi_{\rm pole} - P$ diagram, including a not impossible
boundary for death valley (offset $\delta = 0.8 R_*$, plus inverse Compton
emission with internal reheating controlling the surface temperature at
ages greater than $10^6$ years), the usual Hubble and 
spin-up lines, and the apparanebt strong polar field boundary 
$B = 4.4 \times 10^{13}$ Gauss. \label{fig:phi-P}}
\end{figure}

Satisfactory agreement with the observations occurs with the enhancement of
polar acceleration discovered by Muslimov and Tsygan, but still requires
introducing a special kind of magnetic anomaly (an offset dipole). This
kind of anomaly, however, {\it is} consistent
with the evidence adduced for {\it low altitude} dipolar structure in the
magnetic fields of rotation powered pulsars. If temperatures decline
slower than exponentially with age, resonant Compton scattering
eases the magnitude of the required offset, and creates a gap between 
the observed edge of the pulsar distribution
and the theoretical boundary of death valley, as is shown in Figure 
\ref{fig:phi-P}

\section{Positronium Creation in Strongly Magnetized Neutron Stars?}

In addition to the edge of death valley shown in Figure \ref{fig:phi-P}, 
I have included
a line of constant polar magnetic field equal to the ``critical'' field,
$B_q = 4.4 \times 10^{13}$ Gauss, corresponding to magnetic moments
exceeding $\mu_q \sim 2 \times 10^{31}$ cgs.  The apparent absence of pulsars 
with magnetic moments exceeding $\mu_q$ cannot simply be due to rapid spin
down and therefore a paucity of observable objects because of this simple
evolutionary effect - the empty region would have included stars with ages
in excess of $10^5$ years, if such strong field objects exist. One solution
(the conventional one) is to assume, for reasons unknown, that stars with
magnetic moments in excess of $\mu_q$ simply never form. Another intriguing
possibility is that pair creation is suppressed in such strong fields.  Positronium
formation (Usov and Melrose 1996) is a likely candidate (Arons 1996).  
Provided the bound pairs are not ionized close to the star (photoionization is 
the most likely possibility), the pairs form a neutral gas, not a 
quasi-neutral plasma, which would suppress the low altitude radio emission 
while still providing the outflow needed to power plerions which appear 
to be driven by strong field pulsars but don't show any radiative sign of 
a central compact source (Helfand, Becker and White 1995). Photon
splitting (Baring and Harding 1998) is another means of preventing photons from
converting to pairs in strong fields, thus suppressing radio emission. 
However, as the photons with degraded energy propagate into regions of 
weaker field above the surface, they will convert to pairs as if the
star had had a smaller magnetic moment in the first place. 
\section{Conclusion}
I have shown that polar pair creation based on acceleration
of a steadily flowing, space charged limited non-neutral beam in a
locally dipolar magnetic geometry at low altitude
is consistent with pulsar radio
emission throughout the $P - \dot{P}$ diagram, provided 1) the effect of dragging
of inertial frames is included in estimates of the starvation electric field;
2) the dipole center is strongly offset from the stellar center, perhaps
as much as $0.7-0.8 R_*$; and 3) inverse Compton emission of thermal
photons from a neutron star cooling slower than exponentially at ages
in excess of $10^6$ years plays an important role in the emission of 
magnetically convertible gamma rays.  The development of new diagnostics
of the low altitude magneic field, and gamma ray observations sensitive
to low altitude emission, will eventually provide tests of these ideas.

\section{Acknowledgments}

My research on pulsars is supported in part by NSF grant AST 9528271 and
NASA grant NAG 5-3073, and in part by the generosity of California's taxpayers.

\section{References}
\re Arons, J. 1979, Space Sci. Rev., 24, 437
\re \rule{10mm}{0.1mm}. 1992, {\it Proc. IAU Colloq. No. 128, `The Magnetospheric Structure
  and Emission Mechanisms of Raio Pulsars'}, T.H. Hankins, J.M. Rankin and
  J. A. Gil, eds. (Zielona Gora: Pedagogical University Press), 56
\re \rule{10mm}{0.1mm}. 1993, Ap.J., 408, 160
\re \rule{10mm}{0.1mm}. 1996, {\it Proc. IAU Colloq. No. 160, `Pulsars: Problems and Progress'},
  S. Johnston, M.A. Walker and M. Bailes, eds (San Francisco: Astronomical
  Society of the Pacific), 177
\re Arons, J., and Scharlemann, E.T. 1979, Ap.J., 231, 854
\re Beskin, V.S. 1990, Pis'ma Ast. Zh., 16, 665 
   (Sov. Ast. - Letters, 16)
\re Baring, M.G., and Harding, A.K. 1998, Proc. 4th Compton Symposium,
    C.D. Dermer and J.D. Kurfess, eds. (New York: AIP) 
\re Barnard, J.J., and Arons, J. 1982, Ap.J., 254, 713
\re Becker, W., and Tr\"{u}mper, J. 1997, A\&A, 326, 682
\re Bjornsson, C.-I. 1996, Ap.J., 471, 321
\re Chen, K., and Ruderman, M.A. 1993, Ap.J., 402, 264
\re Deutsch, A.J. 1955, Ann. Ap., 18, 1
\re Dyson, F.J. 1971, ``Neutron Stars and Pulsars: Fermi Lectures 1970'' 
     (Rome: Acadmia Nazionale dei Lincei), 25-26
\re Helfand, D.J., Becker, R.H., and White, R.L. 1995, Ap.J., 453, 741
\re Goldreich, P. and Julian, W. 1969, Ap.J., 157, 869
\re Gurevich, A.V., and Istomin, Ya.N. 1985, Zh.Eksp.Teor.Fiz., 89,3
    (Soviet Physis - JETP, 62, 1)
\re Groth, E.J. 1975, Ap.J. (Supp.), 29, 453
\re Kaspi, V.M. {\it et al.} 1994, Ap. J. Lett., 422, L83
\re Luo, Q. 1996, Ap.J., 468, 338
\re Lyne, A.G., and Manchester, R.N. 1988, MNRAS, 234, 477
\re Mestel, L., and Shibata, S. 1994, 271, 621
\re Meszaros, P. 1992, ``High Energy Raiation from Magnetized Neutron
     Stars'' (Chicago: University of Chicago Press)
\re Muslimov, A., and Tsygan, A.I. 1990, Ast. Zh., 67, 263
    (Soviet Ast., 34, 133)
\re \rule{10mm}{0.1mm}. 1992, MNRAS, 255, 61
\re Narayan, R., and Vivekanand, M. 1982, A\&A, 113, L3
\re Pavlov, G.G., and Zavlin, V.E. 1997, Ap.J.Lett., 490, L91
\re Rankin, J.M. 1990, Ap.J., 352, 247
\re Ruderman, M.A., and Sutherland, P.G. 1975, Ap.J., 196, 51
\re Scharlemann, E.T., Arons, J., and Fawley, W.M. 1978, Ap.J., 222, 297
\re Sturner, S.J., Dermer, C.D., and Michel, F.C. 1995, Ap.J., 445, 736
\re Sturrock, P.A. 1971, Ap.J., 164, 529
\re Tademaru, E. 1974, Ap. and Space Sci., 30, 179
\re Umeda, H., {\it et al.} 1993, Ap.J., 408, 186
\re Usov, V.V., and Melrose, D.B. 1996, Ap.J., 464, 306

\end{document}